\documentclass[12pt]{article}
\pdfoutput=1

\usepackage{amsmath,amsfonts,amssymb,graphicx,psfrag,rotating}
\usepackage[colorlinks=true,urlcolor=blue,citecolor=black,linkcolor=blue]{hyperref}
\usepackage{setspace}
\usepackage{dcolumn}
\usepackage[superscript,biblabel]{cite}

\usepackage{comment}
\usepackage{multirow}
\usepackage{lscape}
\usepackage{bbm}
\usepackage{scalefnt}
\usepackage[autolanguage]{numprint}
\usepackage{booktabs}
\usepackage{bm}
\usepackage{threeparttable}
\usepackage{rotating}
\usepackage{color}
\usepackage{setspace}
\usepackage{sectsty}

\usepackage{caption}
\usepackage{subcaption}

\usepackage{color}
\definecolor{qz}{RGB}{255,0,0}

\usepackage{lmodern}
\fontfamily{lmtt}\selectfont
\usepackage[T1]{fontenc}

\usepackage[top=1in, left=1in, right=1in,
bottom=1in]{geometry}

\makeatletter
\renewcommand{\section}{\@startsection{section}{1}{0em}{\baselineskip}{0.5\baselineskip}{\large\bfseries\large}}
\renewcommand{\subsection}{\@startsection{subsection}{0}{0em}{\baselineskip}{0.5\baselineskip}{\normalfont\bfseries\normalsize}}
\makeatletter

\newcolumntype{.}{D{.}{.}{-1}}
\newcolumntype{d}[1]{D{.}{.}{#1}}

\begin{document}

%%%%%%%%%%%%%%%%%%%%%%%%%%%%%%%%%%%%%%%%%%%%%%%%%%%%%%%%%%%%
%%%%%%%%%%%%%%%%%%%%%% Start of Title Page %%%%%%%%%%%%%%%%%
%%%%%%%%%%%%%%%%%%%%%%%%%%%%%%%%%%%%%%%%%%%%%%%%%%%%%%%%%%%%

%
% TCIMACRO{\TeXButton{Section}{\sectionfont{\bfseries\large\sffamily}}}%
% BeginExpansion
\sectionfont{\bfseries\large\sffamily}%
% EndExpansion
%

% TCIMACRO{\TeXButton{Subsection}{\subsectionfont{\bfseries\sffamily\normalsize
%:
% }}}%
%   BeginExpansion
\subsectionfont{\bfseries\sffamily\normalsize}%
% EndExpansion
%

% TCIMACRO{\TeXButton{noindent}{\noindent}}%
% BeginExpansion
\noindent
% EndExpansion%
% TCIMACRO{\TeXButton{title}{{\sffamily\bfseries\Large
% Cross-screening in observational studies that test many hypotheses}}}%
% BeginExpansion
\noindent\textsf{Title: Comparing the Performance of Statistical Adjustment Methods By Recovering the Experimental Benchmark from the REFLUX trial.}%
% EndExpansion
%

\vspace{1cm}

% TCIMACRO{\TeXButton{noindent}{\noindent}}%
% BeginExpansion
\noindent
% EndExpansion
\textsf{Authors: Luke Keele, Stephen O'Neill and Richard Grieve}%

\vspace{1cm}

% TCIMACRO{\TeXButton{noindent}{\noindent}}%
% BeginExpansion
\noindent
% EndExpansion
\textsf{Affiliation: LSHTM, National University of Ireland Galway, and University of
  Pennsylvania}%

% TCIMACRO{\TeXButton{noindent}{\noindent}}%
% BeginExpansion
\noindent
% EndExpansion

\vspace{1cm}

% TCIMACRO{\TeXButton{noindent}{\noindent}}%
% BeginExpansion
\noindent
% EndExpansion
\textsf{Correspondence address: Luke Keele, Associate Professor, 
  University of Pennsylvania, Email: luke.keele@gmail.com}%

% TCIMACRO{\TeXButton{noindent}{\noindent}}%
% BeginExpansion
%\noindent
% EndExpansion

%\vspace{1cm}

%\noindent\textsf{Suggested running title: XXX}

%\vspace{1cm}

%\noindent\textsf{Type of manuscript: XXX}

%\vspace{1cm}

%\noindent\textsf{Conflict of interest: None.}

%\vspace{1cm}

%\noindent\textsf{Sources of financial support: }

%\vspace{1cm}

%\noindent\textsf{Replication of results: Code for the replication of the results is available as part of an online dataverse and can be retrieved at: http:xxx.xxx.xxx. }

%\vspace{1cm}

%\noindent\textsf{Total words: XXX words.}

%\newpage

%%%%%%%%%%%%%%%%%%%%%%%%%%%%%%%%%%%%%%%%%%%%%%%%%%%%%%%%%%%%
%%%%%%%%%%%%%%%%%%%%%%% End of Title Page %%%%%%%%%%%%%%%%%%
%%%%%%%%%%%%%%%%%%%%%%%%%%%%%%%%%%%%%%%%%%%%%%%%%%%%%%%%%%%%

%\title{Comparing Statistical Methods of Statistical Adjustment}

%\maketitle

\pagestyle{plain}

\begin{abstract}

Much evidence in comparative effectiveness research is based on observational studies. Researchers who conduct observational studies typically assume that there are no unobservable differences between the treated and control groups. Treatment effects are estimated after adjusting for observed differences between treated and controls. However, treatment effect estimates may be biased due to model misspecification.  That is, if the method of treatment effect estimation imposes unduly strong functional form assumptions, treatment effect estimates may be significantly biased. In this study, we compare the performance of a wide variety of treatment effect estimation methods. We do so within the context of the REFLUX study from the UK. In REFLUX, after study qualification, participants were enrolled in either a randomized trial arm or patient preference arm.  In the randomized trial, patients were randomly assigned to either surgery or medical management. In the patient preference arm, participants selected to either have surgery or medical management. We attempt to recover the treatment effect estimate from the randomized trial arm using the data from the patient preference arm of the study. We vary the method of treatment effect estimation and record which methods are successful and which are not. We apply over 20 different methods including standard regression models as well as advanced machine learning methods. We find that simple propensity score matching methods perform the worst. We also find significant variation in performance across methods. The wide variation in performance suggests analysts should use multiple methods of estimation as a robustness check.

\end{abstract}
%keelehsreditorial
\doublespacing

\clearpage

\section{Introduction}
\label{sec:intro}

Comparative effectiveness research focuses on evaluating the effects of health care strategies on patient outcomes and includes the evaluation of medical strategies, prevention methods, diagnostic testing, devices, surgeries and rehabilitative techniques. CER also includes the evaluation of innovations in health care delivery, organization and financing, as well as public health interventions such as vaccinations. The primary scientific aim of CER is to quantify the causal effect of an intervention on patient outcomes \cite{hernan2018c}. To learn about effectiveness, one must also consider what \emph{would have happened} under different circumstances (e.g., if those treated had taken control). Causal inference in statistics focuses on formulating these questions mathematically, exploring whether answers can be gleaned from data, and if so, determining how well and with what statistical methods. 
  
As such, research designs and statistical methods for causal inference form a key part of CER. While randomized trials can rule out bias from unobserved confounders, results from observational studies are also a critical part of the evidence base in CER. An observational study is an empirical comparison of treated and control groups where the objective is to elucidate cause-and-effect relationships in contexts where it is not feasible to use controlled experimentation and subjects select their own treatment status \cite{Cochran:1965}. In an observational study of treatment effects, analysts use pretreatment covariates and a statistical adjustment strategy to remove overt differences in the treated and control groups. Estimated treatment effects from an observational study can suffer from bias primarily from two sources. The first source of bias is from unmeasured confounders. In an observational study, the investigator must assume there are no unobservable differences between the treated and control groups. This assumption is often called ``selection on observables,'' since the analyst asserts that there is some set of covariates such that treatment assignment is random conditional on these covariates \cite{Barnow:1980}. Critically, this assumption cannot be tested with observed data \cite{Manski:2007}. 

However, even if the selection on observables holds, much can go wrong, since model misspecification may be a second source of bias. That is, if the analyst selects a method of statistical estimation that lacks flexibility, the treatment effect estimate may be biased due to functional form misspecficiation. For example, functional form misspecification can arise if a continuous control variable is modeled as linear when it in fact has a nonlinear association with the outcome. In general, it would be a tragedy to bias a treatment effect estimate if one is fortunate enough to be in a situation with no unmeasured confounders.  Due to this concern, a large number of flexible estimation methods for treatment effects have been developed. Here, we conduct an investigation into the comparative performance of a wide range of estimation methods for observational study designs. Our research is built on an innovative study design that allows us to benchmark performance against results from a randomized experiment. We review this study design next.

\section{Study Design: RCT Benchmarking}

Evaluations of statistical estimation methods for treatment effects in observational studies typically use two different methods. The most common method is statistical simulation. Simulations allow for precise descriptions of statistical performance, since the truth is known. However, simulation scenarios require abstracting away from the often messy nature of real applications. One alternative to simulation is RCT benchmarking. In an RCT benchmark design, the investigator conducts an observational study that is focused on recovering treatment effect estimates from a randomized trial. That is, the analyst creates a scenario where the estimates from an observational study can be directly compared to the estimates from a randomized trial. We deploy this strategy here. Under this design, data from an RCT are used to estimate the benchmark treatment effect. In the next phase, data from an observational study are used to estimate a treatment effect that can be compared to the RCT estimate. The observational study part of the design can be conducted in two ways. First, one can remove the treatment group from the experimental study and replace it with a control group from a new data source. This design was used in the well-known Lalonde study \cite{Lalonde:1986}. Alternatively, the observational study component is based on an entirely new data source where some units are exposed to the same treatment studied in the RCT \cite{fralick2018use}. Our study follows the second template. Next, we describe the study protocol more specifically.

\subsection{REFLUX Study Design}

We employ data from a study conducted in the United Kingdom (UK). Here, we review that study and outline how we use it in our investigation. Gastro-Oseophageal Reflux Disease (GORD) develops when reflux of the stomach contents cause troublesome symptoms or complications which adversely affect a patients' well-being. Standard medical management for patients with GORD is via Proton Pump Inhibitors (PPIs) to suppress acid reflux. While PPIs are effective, there is the concern that long-term acid suppression from PPIs may be associated with increased risk of chronic hypergastrinaemia and gastric cancer. One alternative treatment to PPIs is laparoscopic surgery. However, surgical management, while minimally invasive, carries risk of side effects, such as infection.

To compare the effectiveness of these two treatment plans, investigators conducted REFLUX, a multicenter RCT, designed to estimate the effectiveness of laparoscopic surgery for patients with moderately severe GORD. In the study, patients were randomly assigned to medical management or laparoscopic surgery \cite{grant2008effectiveness,grant2013clinical}. REFLUX was not just an RCT, however, as it also included a concurrent parallel patient preference arm \cite{grant2008effectiveness}. Both arms of the study included patients from the same 21 hospitals in the UK over the same time period (2001-4). All study participants were asked whether they would consent to participate in the RCT, and those who agreed were randomly assigned to either medical management or surgery. The patients who chose not to participate in the RCT were asked to join the study in the patient preference arm. These patients were then treated according to their preference for either medical management or surgery.

Both arms of the study used the same pragmatic design, and recruited patients according to minimal inclusion criteria including: the patient was already having medical management with PPI, and the recruiting clinician was uncertain as to whether surgery or medical management would be more beneficial. Ineligible patients or those who did not consent to participate were excluded. In both arms of the study, patients were asked to complete a baseline questionnaire which included disease-specific and generic measures of health status. We focus on two primary endpoints. The first endpoint is EQ-5D, a standardized measure of health status developed by the EuroQol Group. We focused on EQ-5D measured at 3 months after surgery. We use this outcome measured at 3 months to minimize the effects of missing outcome data. The second endpoint we use is a quality of life measure focused specifically on reflux symptoms. The study enrolled 810 participants. A total of 357 patients were enrolled into the RCT (179 assigned to medicine, 178 to surgery) and 453 in the preference study (192 selected medicine, 261 selected surgery).  

One strength of the REFLUX study is that it meets recently developed criteria to ensure that RCT and non-randomized study (NRS) results are comparable \cite{lodi2019effect}.  More specifically, as part of REFLUX the two arms used a common eligibility criteria and treatment strategy.  Moreover, outcome measurement and follow-up periods were identical for both arms of REFLUX. This harmonization ensures that differences between the RCT and NRS estimates are not a result of differing study protocols. 

We use the REFLUX study for RCT benchmarking in the following way. First, we use the patients enrolled in the RCT arm to estimate the experimental benchmark for the two outcome measures. We then seek to estimate the same effect using the patient preference arm. We estimate the effect of surgery by comparing those who selected to have surgery to those who selected medical management. To estimate the effect within the non-randomized study (NRS), we control for baseline covariates to make these two groups comparable. We then compare the estimate from the NRS to the RCT and test whether they different significantly. In our study, we seek to understand whether the difference between the NRS and RCT estimates is a function of the treatment effect estimation method we use. Next, we review approaches to and estimation methods for treatment effect estimates in a NRS.

\section{Approaches to Statistical Adjustment for Observed Confounders}

In our study, we adjust for observed confounders in the observational arm of the design and test whether the NRS estimates are close to the estimates from the RCT.  As we noted above, analysts can apply a wide variety of statistical methods to adjust for observed confounders. In our study, we employ many different statistical adjustment methods. Our goal is to understand whether some methods of statistical adjustment perform better in this setting than others. Here, we outline the three broad approaches to statistical adjustment that we will compare. First, we define some notation and the necessary identification assumptions needed in the NRS component of REFLUX. 

We assume patients in the REFLUX NRS (indexed by $i=1,\dots,n$) form the population, that the treatment is binary ($Z_{i} = 1$ (surgery), $Z_{i} = 0$ (control), and we observe an outcome variable $Y_i$. Next, we outline the potential outcomes framework \cite{Neyman:1923a,Rubin:1974}. Prior to treatment, each patient has two potential responses: $(Y_{i}(1), Y_{i}(0))$. The outcomes that we actually observe are a function of potential outcomes and treatment assignment: $Y_{i} = Z_{i}Y_{i}(1) + (1 - Z_{i})Y_{i}(0)$. We also have a matrix of observed, pretreatment covariates, $\mathbf{X}_{i}$. For each patient, there is possibly an unobserved covariate $u_{i}$. Our notation implicitly assumes that the stable unit treatment value assumption (SUTVA) holds \cite{Rubin:1986}. SUTVA is comprised of the two following components: 1) the treatment levels of $Z$ (1 and 0) adequately represent all versions of the treatment and 2) a subject's outcomes are not affected by other subjects' exposures.  The first component of SUTVA is often referred to as the consistency assumption in the epidemiology literature. The second component of SUTVA rules out a subject's outcomes being affected by other subjects' exposures. In general, it is difficult to conceive of how the selection of surgical versus non-surgical care for one patient could affect outcomes for other patients even within the same hospital. 

In an NRS, we also invoke additional assumptions that we describe as ``selection on observables.'' Under the selection on observables assumption, the analyst asserts that treatment assignment is random conditional on a set of observed covariates \cite{Barnow:1980}. Formally we assume that treatment assignment only depends on observed data:
\[
Pr(Z_{j} = 1|Y_{i}(1), Y_{i}(0), \mathbf{X}_{i}, u_{i}) = Pr(Z_{i} = 1| \mathbf{X}_{i}).
\]
\noindent and the probability of treatment is strictly greater than zero and less than one over the support of $Z_i$:
\[
0 < Pr(Z_{i}) < 1
\]
\noindent Under these two assumptions, there are three broad approaches to statistical adjustment.

\subsection{Outcome Focused Methods}

The dominant approach to statistical adjustment focuses on modeling the outcome. Here, the analysts builds a statistical model for the conditional expectation of $Y_i$ given $Z_i$ and $X_i$: $E(Y_i|Z_i, X_i)$. The goal is to estimate the causal effect of $Z_i$ on $Y_i$ while controlling for observed confounders. Using a model for the outcome has long been the dominant approach and has traditionally employed methods such as linear or logistic regression. Though as we outline below, a wide variety of more flexible estimation methods can be employed. 

\subsection{Treatment Focused Methods}

The primary alternative to outcome modeling is modeling the treatment assignment process. That is, if the true probability of treatment (the propensity score) is known, one can use the propensity score to estimate treatment effects \cite{Rosenbaum:1983}. Since, the true propensity score is typically unknown, the investigator estimates a model for the propensity score, which is a model for the conditional expectation of $Z_i$ given $X_i$: $E(Z_i|X_i)$. The estimated propensity score can be used with either weighting or matching estimators for treatment effect estimation. Alternatively, many matching methods do not use the propensity score but instead use a direct measure of covariate distance such as the Mahalanobis distance \cite{Mahalanobis:1936,Rubin:1980}. However, these matching methods also model $E(Z_i|X_i)$ instead of $E(Y_i|Z_i, X_i)$. Critically, under this approach, statistical adjustment is done without reference to outcomes.

\subsection{Outcome--Treatment Focused Methods}

More recently, a class of statistical adjustment methods have been developed which model both the outcome and treatment. Under this approach, a model for the treatment assignment and two outcome models are combined to obtain a single treatment effect estimate. This approach is often referred to as doubly robust, since it will consistently estimate the treatment effect when either the propensity score model is correctly specified or the outcome models are correctly specified \cite{Scharfstein:1999a}. 

\subsection{Estimation Methods}

For each of the three broad approaches outlined above, estimation methods are required for one or more conditional expectation functions. That is, typically, one is either estimating a model for the outcome, treatment, or both. A very large number of methods can be used to estimate these models. As we noted above, the simplest approach is to use either linear or logistic regression. However, these models are based on simple functional forms that impose strong linearity and additivity assumptions. The danger is that bias from functional form misspecification may be significant. Alternatively, a more flexible model such as a generalized additive model or kernel regression may also be used. The general trend, however, has been to employ increasingly flexible models usually taken from the literature on statistical prediction and widely referred to as ``machine-learning'' (ML) methods. For example, McCaffrey et al\cite{Mccaffrey:2004} used generalized boosting to flexibly model the propensity score, while Hill \cite{Hill:2011} proposes using bayesian additive regression trees (BART) to flexibly model the outcome. Alternatively, double robust approaches may use a single ML method such as random forests \cite{wager2017estimation} or combine several \cite{sinisi2007super} for treatment effect estimation. While most methods of adjustment rely on a regression model, matching and weighting estimators that use covariate distance instead of a propensity score distance, do not estimate a regression model in the traditional sense. Covariate distance is minimized using an optimization method. This approach avoids the use of a regression model. Once the matching or weighting is complete, a simple nonparametric estimator such as the signed rank tests may then be employed to estimate the treatment effect. However, a regression based model may also be used to estimate the treatment effects after matching or using weights \cite{imbens2015matching}. 

As we outlined above, we vary the approach to statistical adjustment in the NRS arm of the REFLUX study. To that end, we use all three approaches: outcome, treatment, or double robust.  We also vary the estimation methods used for each of the approaches. While, there are an almost endless number of methods we could employ, we focused on the most prominent approaches, but also let available software options inform our choices. Broadly, we selected a set of ML methods and then a set of methods that are widely available in Stata or R.

First, we selected three highly flexible machine learning methods. Specifically, we selected Bayesian additive regression trees. BART was originally designed for statistical prediction \cite{chipman2010bart}, but is an example of a ML method being used for statistical adjustment under selection on observables \cite{Hill:2011a}. We use the version of BART that only models outcomes. Recent work has incorporated treatment assignment models into BART \cite{hahn2017bayesian}. The second ML method we selected was generalized random forests (GRF) \cite{2016arXiv161001271A}. While GRF is similar to BART in that it implements a flexible model for the outcome, under GRF there is also a flexible model for the treatment assignment process; as such it is doubly robust. Finally, we also selected a method that uses a Super Learner (SL) combined with Targeted Maximum Likelihood Estimation (TMLE) \cite{sinisi2007super,van2011targeted,gruber2015ensemble}. Under SL, the analyst selects among a set of prediction methods---learners---that will then be combined in an ensemble. The set of learners selected by the investigator are used to make out-of-sample predictions through cross-validation. The predictions from each learner are combined according to weights that minimize the squared-error loss from predictions to observations. These weights are then used to combine the fitted values from each learner when fit to the complete data. Then TMLE is applied to produce an estimate of the ATE or ATT. We used three different learners: (1) random forests, (2) lasso, and (3) generalized boosting. 

While statistical adjustment methods based on ML have received considerable attention as of late \cite{dorie2017automated}, simpler methods are much more commonly used. To that end, we also applied several other methods for statistical adjustment. In selecting additional methods for comparison, we used software availability to guide our choices. That is, we first used the options made available in Stata. In Stata, one can estimate treatment effects via either matching or weighting using the teffects command. The teffects command allows for four different forms of statistical adjustment via: (1) regression adjustment, (2) propensity score weighting, (3) augmented propensity score weighting, (4) propensity score weighting and regression adjustment. The last category is a form of a double robust estimator. Stata also includes matching estimators. We used the Stata commands for propensity score matching (teffects psmatch) and nearest neighbor matching (teffects nnmatch). Finally, Stata includes an option to match with additional regression adjustment after matching. We added this method of adjustment as well.

Next, we included a set of statistical adjustment methods that are available in R. In R, we estimated treatment effects via regression adjustment, propensity score weighting, augmented propensity score weighting, and a doubly robust estimator based on propensity score weighting. However, R also includes a wide variety of matching estimators that are unavailable in Stata. To that end, we included three different forms of matching. First, we included a standard match on the propensity score using an optimal form of matching \cite{Rosenbaum:1989}. Next, following the guidelines in \cite{Rosenbaum:2002} we matched using the Mahalanobis distance with a propensity score caliper. Finally, we used a more advanced method of matching known as cardinality matching \cite{Zubizarreta:2014}. Cardinality matching seeks to target a specified set of balance constraints and will trim the sample to meet those balance targets. In a recent comparison of statistical methods via simulation, cardinality matching had the best properties relative to a variety of matching estimators \cite{resa2016evaluation}. For all three matching methods, we estimated treatment effects in two ways. First, we estimated the treatment effect as the average difference in matched pair outcomes. Second, we estimated the treatment effect via a regression model using the matched data, but also included additional covariates for bias correction.  

In sum, we use a variety of statistical methods for estimating treatment effects within the context of REFLUX.  The methods we use are based on all three approaches and include simple methods such as regression to more advanced methods based on machine learning. We seek to draw conclusions about whether more complex methods appear to produce treatment effect estimates that are closer to the experimental benchmark in REFLUX. The primary strength of this research design is that it allows us to evaluate statistical adjustment methods in a realistic setting.

% STATA
% reg adjust
% ipw
% aipw
% DR
% PS match
% NN match
% NN match + RA

%R
% Reg
% IPW
% AIPW
% DR
% PS score
% PS score + RA
% Cov Match
% Cov Match + RA
% Card Match 
% Card Match + RA

%Zubizarreta:2014
%resa2016evaluation

\section{Data Analysis}

Next, we harmonized the data analysis between the RCT arm and observational study arms of REFLUX, to further ensure that the estimates are comparable \cite{lodi2019effect}. This harmonization is critical, since there was noncompliance with the trial protocol in the RCT arm. Specifically, in the RCT arm, of the 178 patients that were assigned to surgery 67 received medical management; whereas in the medical management arm, ten patients crossed over to surgery. 

However, in the NRS arm of REFLUX a similar pattern occurred. While patients initially selected their course of treatment, due to feedback from clinicians, some changed their course of treatment. In the NRS arm, while 261 selected surgical treatment, 43 patients later changed to medical management. Moreover, of the 192 patients that selected medical management, six later crossed-over and had surgery. Given that in both arms of REFLUX, there was crossover, the correct comparison is that of intention-to-treat (ITT) effects. The ITT estimate, in both arms, is the effect of assignment to treatment rather than exposure to the treatment. While we could seek to estimate the effect of treatment exposure in both the RCT and the NRS that would require additional assumptions.
%exposure. Alternatively, we could estimate the adjusted per-protocol (PP) estimate from the RCT \cite{murray2016adherence}. This would ensure that that we are comparing only exposed patients from the RCT. However, we found the ITT estimate was nearly identical compared to the adjusted PP estimate (0.95 vs 0.98).  Due to the high level of comparability, we use the ITT estimate from the original trial.

In our analysis, we conduct a formal statistical comparison between the ITT estimate from each statistical estimator and the experimental benchmark. More formally, for each method of adjustment, we then estimated the following measure of standardized bias:
\[
\frac{\hat{\tau}_{rct} - \hat{\tau}_{obs}}{SD_{rct=0}}
\]
\noindent where $\hat{\tau}_{rct}$ is the estimate from the original RCT, $\hat{\tau}_{obs}$ is the estimated treatment effect from the NRS, and $SD_{rct=0}$ is the standard deviation from the control group in the RCT.  This is a standard measure for studies of this type \cite{wong2017empirical}. Thus, we report how far the estimates from observational study methods are from the RCT estimate. We also include 95\% confidence intervals for this measure of standardized bias. We also compared methods using mean-squared error (MSE). We calculated MSE as the length of the 95\% confidence interval added to the bias-term squared.

\begin{table}[htbp]
\centering
\begin{threeparttable}
\caption{Balance Table for Patient Preference Arm in REFLUX Study}
\label{tab.bal1}
\begin{tabular}{lcccc}
 \toprule
 & Treated Mean & Control Mean & Std. Diff. & p-value \\
\midrule
Age & 45.21 & 49.04 & -0.32 & 0.00 \\ 
  Female & 0.38 & 0.43 & -0.11 & 0.34 \\ 
  Duration of Symptoms & 27.63 & 27.53 & 0.03 & 0.82 \\ 
  BMI & 59.56 & 41.92 & 0.28 & 0.01 \\ 
  Employment Category 1 & 0.64 & 0.53 & 0.23 & 0.04 \\ 
  Employment Category 2 & 0.16 & 0.10 & 0.15 & 0.17 \\ 
  Employment Category 3 & 0.20 & 0.37 & -0.37 & 0.00 \\ 
  Education Category 1 & 0.56 & 0.51 & 0.10 & 0.34 \\ 
  Education Category 2 & 0.27 & 0.24 & 0.09 & 0.43 \\ 
  Education Category 3 & 0.17 & 0.25 & -0.22 & 0.05 \\ 
  Heartburn Symptoms & 49.08 & 73.96 & -1.08 & 0.00 \\ 
  Gastro Symptoms & 47.76 & 60.75 & -0.59 & 0.00 \\ 
  Nausea Symptoms & 77.42 & 90.20 & -0.77 & 0.00 \\ 
  Reflux Activity Score & 74.47 & 87.14 & -0.88 & 0.00 \\ 
  Gastro Symptoms 1 & 74.80 & 83.84 & -0.45 & 0.00 \\ 
  Health Quality Score & 0.68 & 0.75 & -0.31 & 0.00 \\ 
\bottomrule
\end{tabular}
\begin{tablenotes}[para]
Note: Std. Diff: the difference in means divided by the standard deviation before adjustment.
\end{tablenotes}
\end{threeparttable}
\end{table}

In the NRS, we adjust for a large number of baseline covariates collected for all patients. These covariates include age, sex, BMI, employment status (3 categories), age left education, initial EQ-5D, and 5 scales that measured reflux symptoms including heartburn and nausea. The data also contain indicators for the 21 centers that participated in the study. 

Table~\ref{tab.bal1} contains balance statistics by treatment in the NRS arm of REFLUX.  We find clear differences between the patients that selected surgery versus medical management. For example, surgical patients had notably lower health quality scores and very different scores on the symptom scales. Table~\ref{tab.bal2} compares balance between the NRS and RCT arms of the study. In the RCT, randomization balances observed covariates as expected, while only a few of the covariates are balanced in the NRS.

\begin{table}[htbp]
\centering
\begin{threeparttable}
\caption{Balance Table for Patient Preference Arm in REFLUX Study}
\label{tab.bal2}
\begin{tabular}{lcccc}
 \toprule
 & \multicolumn{2}{c}{RCT} & \multicolumn{2}{c}{NRS}\\
 & Std. Diff. & p-value & Std. Diff. & p-value \\
\midrule
Age & 0.18 & 0.11 & -0.32 & 0.00 \\ 
  Female & 0.06 & 0.58 & -0.11 & 0.34 \\ 
  Duration  of Symptoms & 0.01 & 0.89 & 0.03 & 0.82 \\ 
  BMI & 0.09 & 0.40 & 0.28 & 0.01 \\ 
  Employment Category 1 & 0.11 & 0.33 & 0.23 & 0.04 \\ 
  Employment Category 2 & -0.06 & 0.57 & 0.15 & 0.17 \\ 
  Employment Category 3 & -0.08 & 0.49 & -0.37 & 0.00 \\ 
  Education Category 1 & 0.05 & 0.63 & 0.10 & 0.34 \\ 
  Education Category 2 & -0.04 & 0.74 & 0.09 & 0.43 \\ 
  Education Category 3 & -0.03 & 0.80 & -0.22 & 0.05 \\ 
  Heartburn Symptoms & -0.08 & 0.47 & -1.08 & 0.00 \\ 
  Gastro Symptoms & -0.05 & 0.63 & -0.59 & 0.00 \\ 
  Nausea Symptoms & 0.06 & 0.60 & -0.77 & 0.00 \\ 
  Reflux Activity Score & -0.03 & 0.79 & -0.88 & 0.00 \\ 
  Gastro Symptoms 1 & 0.11 & 0.33 & -0.45 & 0.00 \\ 
  Health Quality Score & 0.01 & 0.94 & -0.31 & 0.00 \\  
\bottomrule
\end{tabular}
\begin{tablenotes}[para]
Note: Std. Diff: the difference in means divided by the standard deviation of control group. NRS: Nonrandomized study.
\end{tablenotes}
\end{threeparttable}
\end{table}

\section{Results}

\begin{figure}
\centering
\begin{subfigure}[b]{.47\textwidth}
  %\centering
  \includegraphics[width=\textwidth]{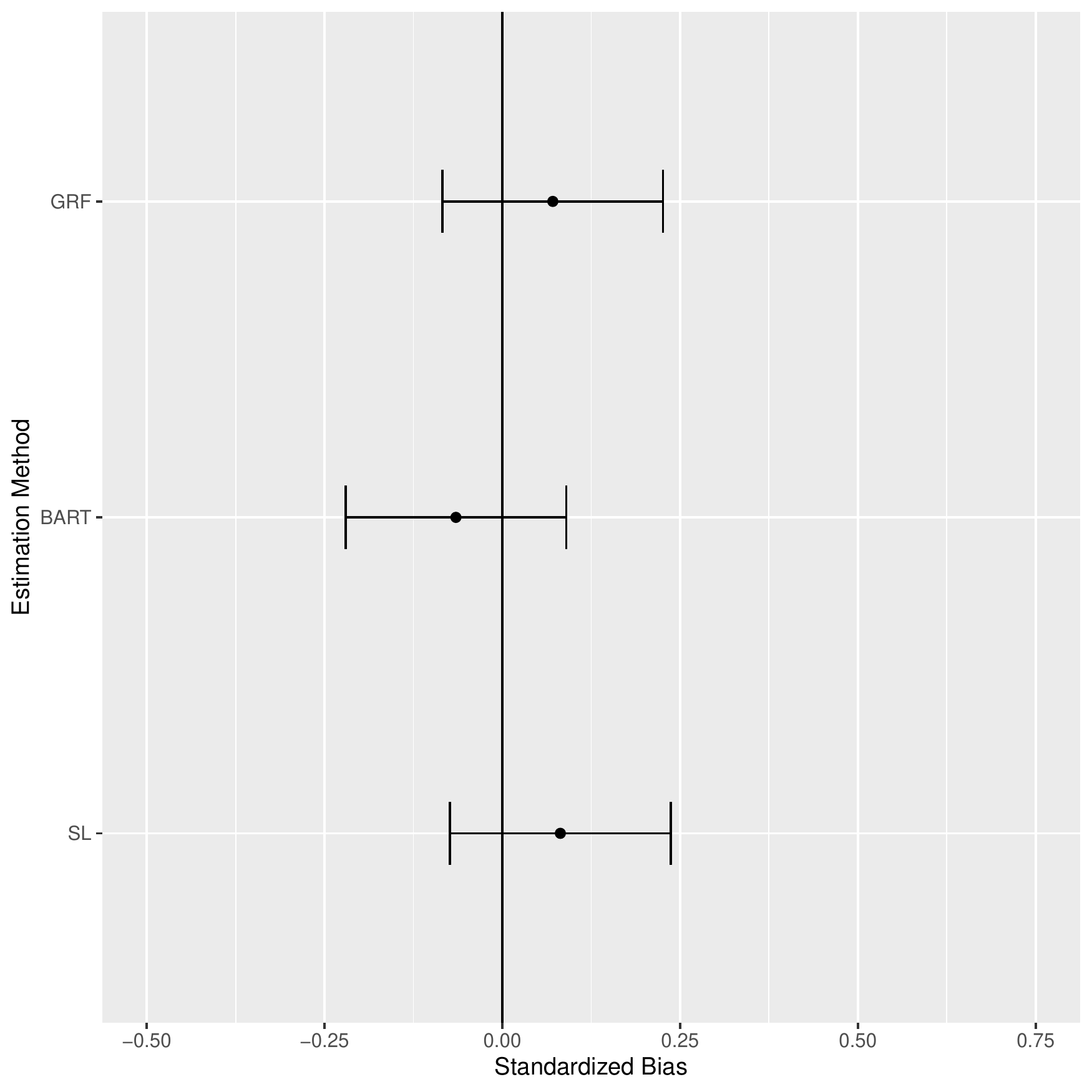}
  \caption{Health Status}
  \label{fig:ml1}
\end{subfigure}
\begin{subfigure}[b]{.47\textwidth}
  %\centering
  \includegraphics[width=\textwidth]{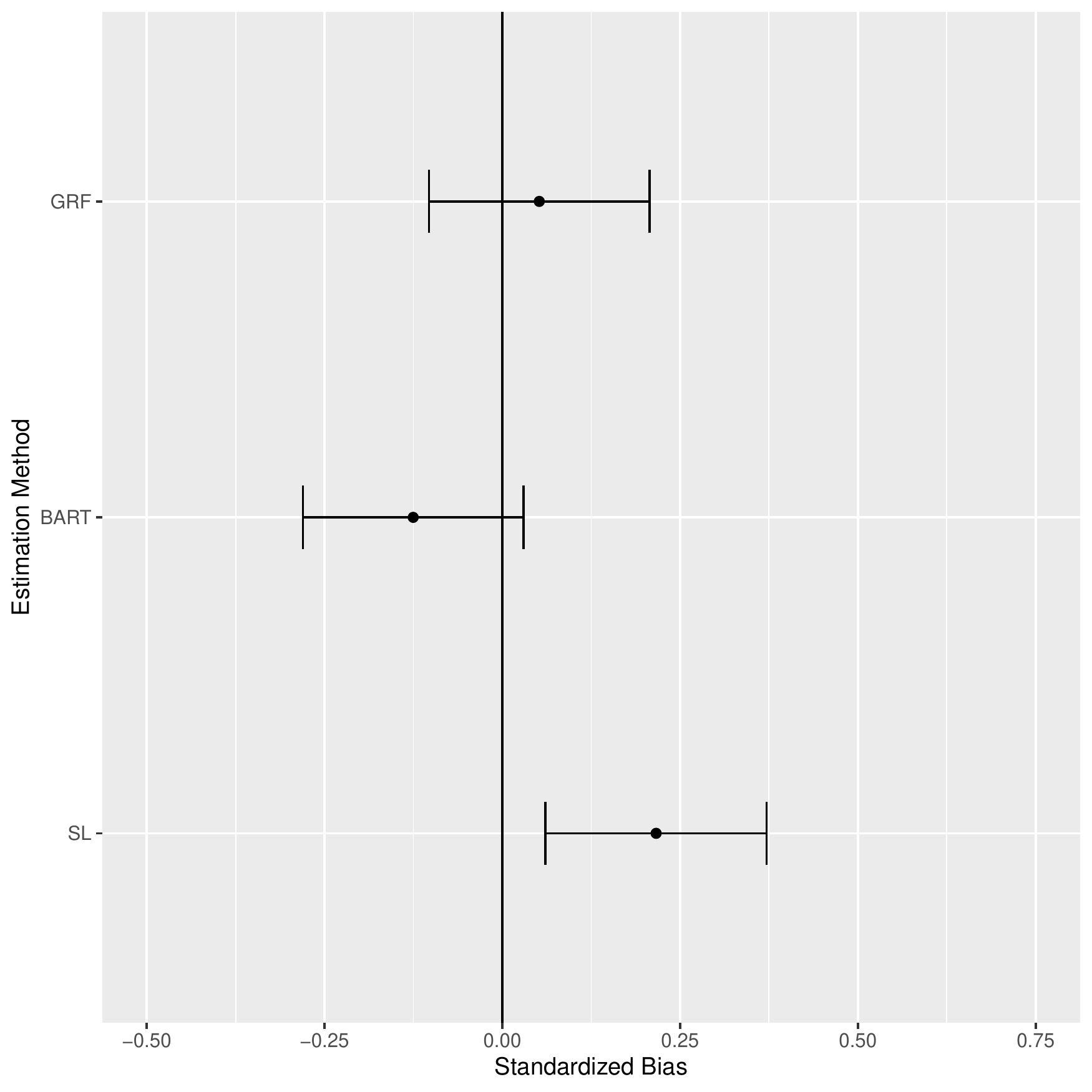}
  \caption{Quality of Life}
  \label{fig:ml2}
\end{subfigure}

\caption{Bias in statistical estimates for on machine learning methods of statistical adjustment. Estimates are standardized difference between NRS and RCT treatment effects.}
\label{fig:ml_plots}
\end{figure}

Next, we present the results from our analysis. Figure~\ref{fig:ml_plots} contains the results for the three ML based methods for estimating treatment effects. For the health status outcome, the confidence intervals cover zero which implies that the estimated effects from the NRS are indistinguishable from the RCT estimate. However, for the quality of life outcome, this is only true for the BART and GRF methods. For the estimate based on TMLE and a Superlearner, the estimated bias exceeds .2 standard deviations and the 95\% confidence interval does not cover zero. To further investigate why this might be the case, we iterated added additional learners to the Superlearner. However, this did not improve the estimates.  Moreover, while the BART confidence interval does overlap zero, the estimated bias is nearly .15 standard deviations, which one might denote as large but not statistically significant.  

Next, we report results based on the methods available in Stata. For the health status outcome, one specific pattern stands out: both matching estimators returns estimates that are significantly different from the RCT estimate. However, once additional bias reduction is added via regression adjustment, the matched estimate is no longer statistically different from the RCT estimate. The source of the bias in these matching methods is straightforward: post-matching balance checks reveal significant imbalances remain. All the estimates based on weighting have confidence intervals that cover zero. While this is also true for the estimate based on standard regression adjustment, this estimate has higher levels of bias than those based on weighting. For the quality of life outcome, only the estimate based on augmented propensity score weighting is close to the RCT estimate.  For every other method, the estimated bias is statistically significant. Here, it is worth noting that additional bias reduction via modeling the outcome after matching or with the doubly robust estimator is ineffective.

\begin{figure}[hbtp]
\centering
\begin{subfigure}[b]{.47\textwidth}
  %\centering
  \includegraphics[width=\textwidth]{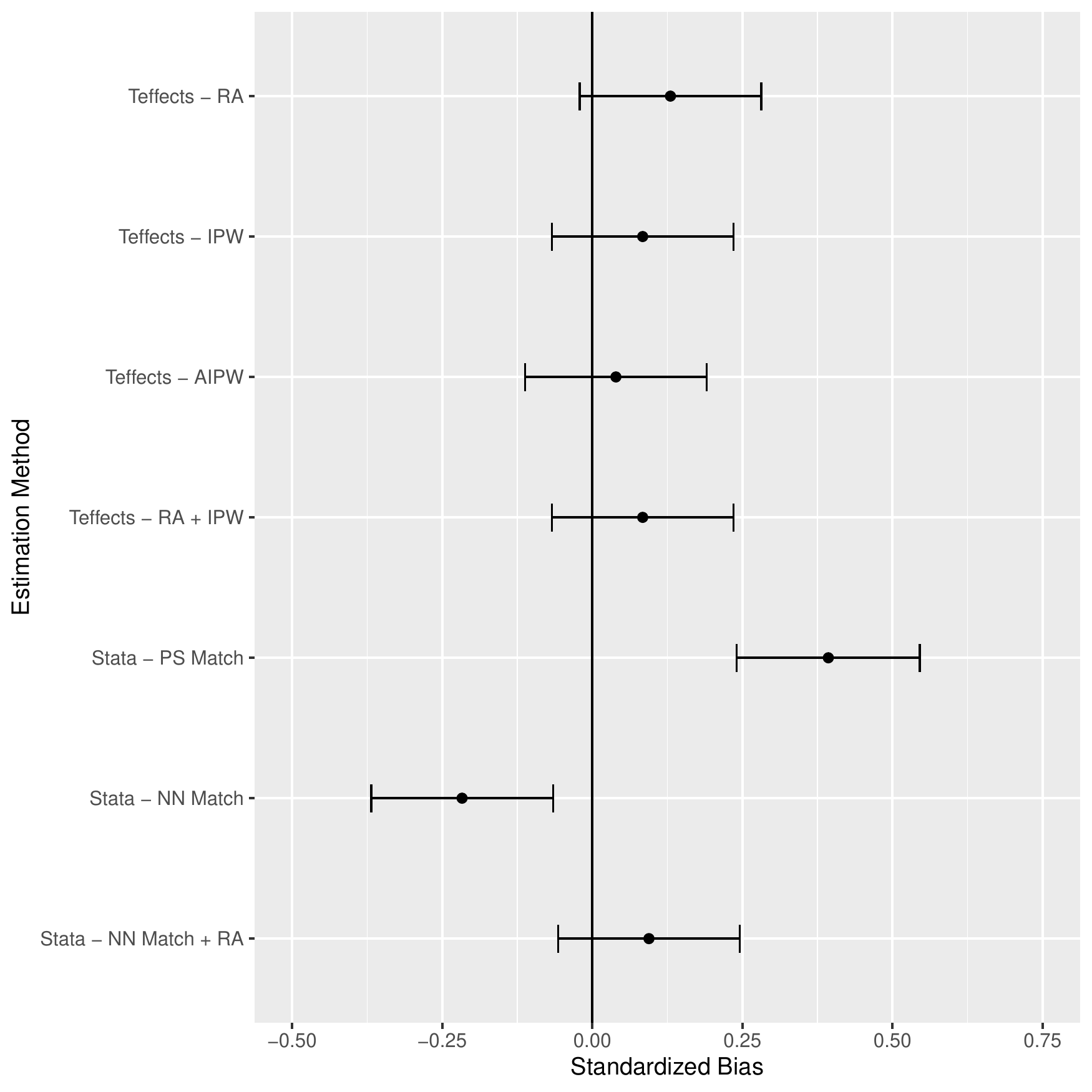}
   \caption{Health Status}
  \label{fig:st1}
\end{subfigure}
\begin{subfigure}[b]{.47\textwidth}
  %\centering
  \includegraphics[width=\textwidth]{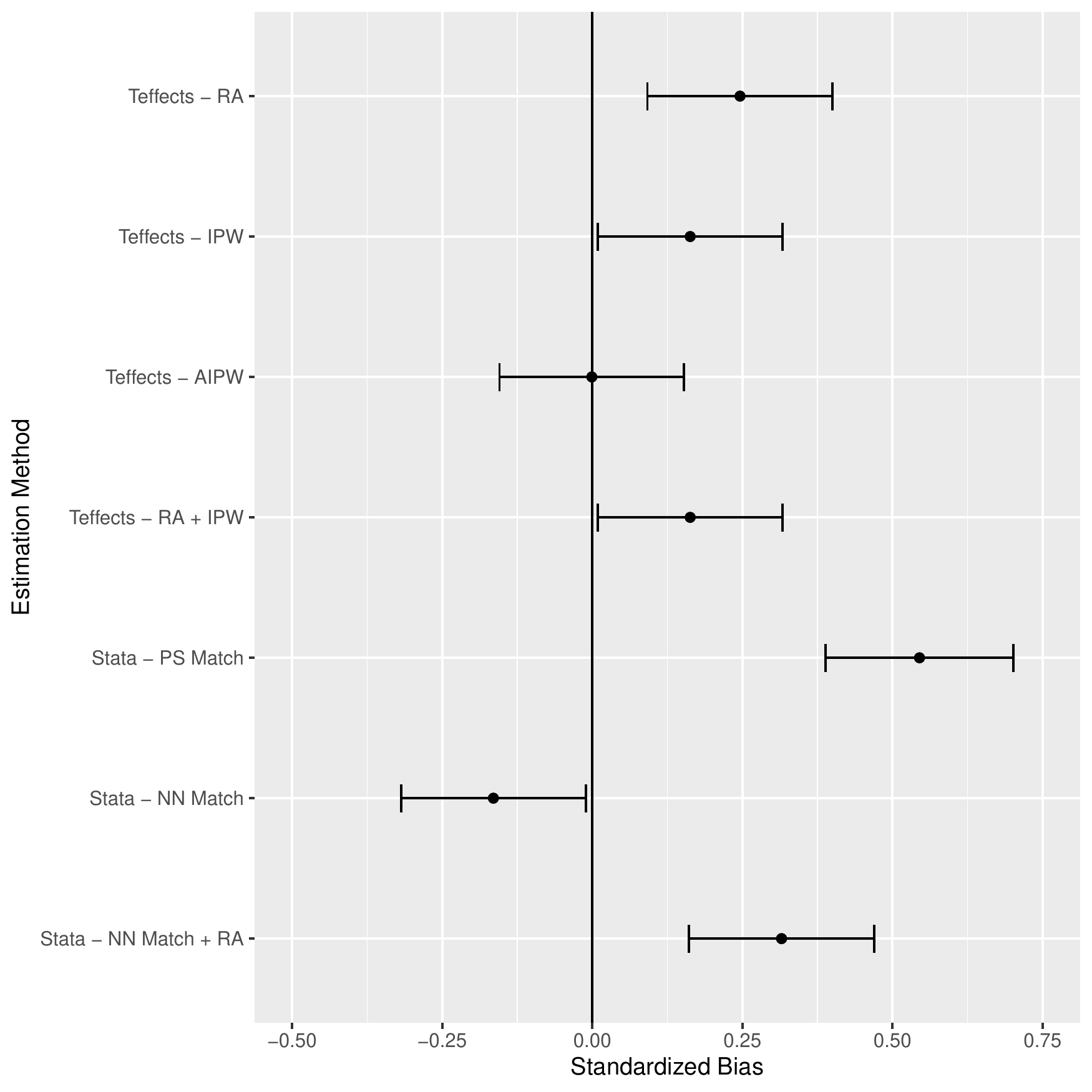}
  \caption{Quality of Life}
  \label{fig:st2}
\end{subfigure}

\caption{Bias in statistical estimates based on methods of statistical adjustment available in Stata. Estimates are standardized difference between NRS and RCT treatment effects}
\label{fig:stata_plots}
\end{figure}

Finally, we report the results using methods in R.  For the health status outcome, only the estimate based on a propensity score match statistically differs from the RCT estimate. Again, we found that the propensity score match produces large imbalances. However, once we include additional bias reduction via a regression model, the propensity score matching estimate displays little bias. In fact, we see across all three matched estimates, adding an outcome model reduces the bias. Also we find that only the doubly robust estimate overestimates the RCT benchmark. For the quality of life outcome, we find that two of the matching estimators produce significant amounts of bias. That is, the estimates from a propensity score and Mahalanobis distance match, differ significantly from the RCT benchmark. Both matching methods failed to balance key covariates. However, in both cases, using a regression model after matching produces significant bias reduction---the estimates are nearly identical to the RCT benchmark. One interesting pattern is apparent for cardinality matching.  Cardinality matching produces little bias even without additional regression adjustment, and additional regression adjustment only produces minimal further bias reduction. Unlike other matching methods, cardinality matching directly targets high levels of balance at the start of the matching process. Finally, all the estimates based on weighting produce results that do not differ significantly from the RCT benchmark.

\begin{figure}[hbtp]
\centering
\begin{subfigure}[b]{.47\textwidth}
  %\centering
  \includegraphics[width=\textwidth]{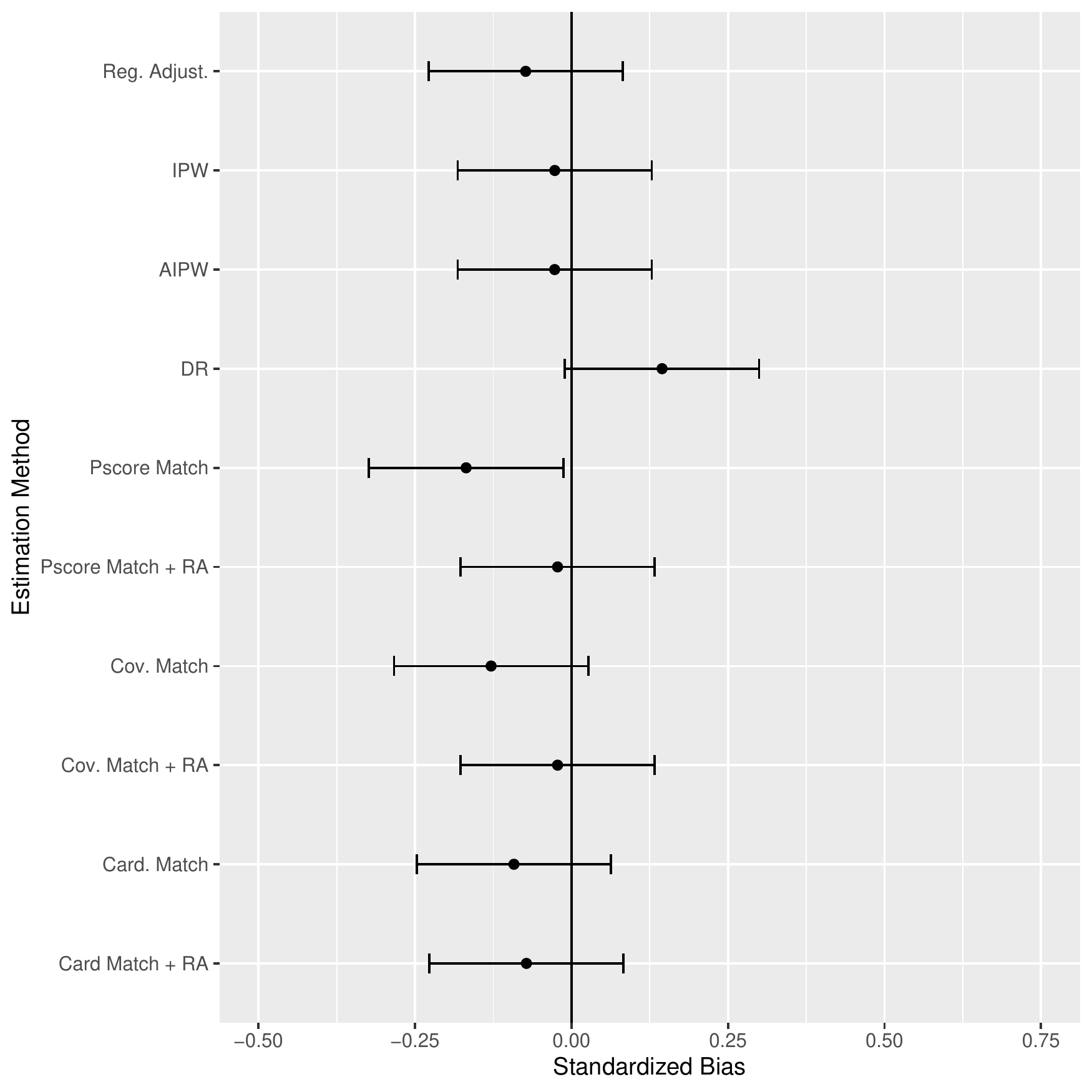}
  \caption{Health Status}
  \label{fig:r1}
\end{subfigure}
\begin{subfigure}[b]{.47\textwidth}
  %\centering
  \includegraphics[width=\textwidth]{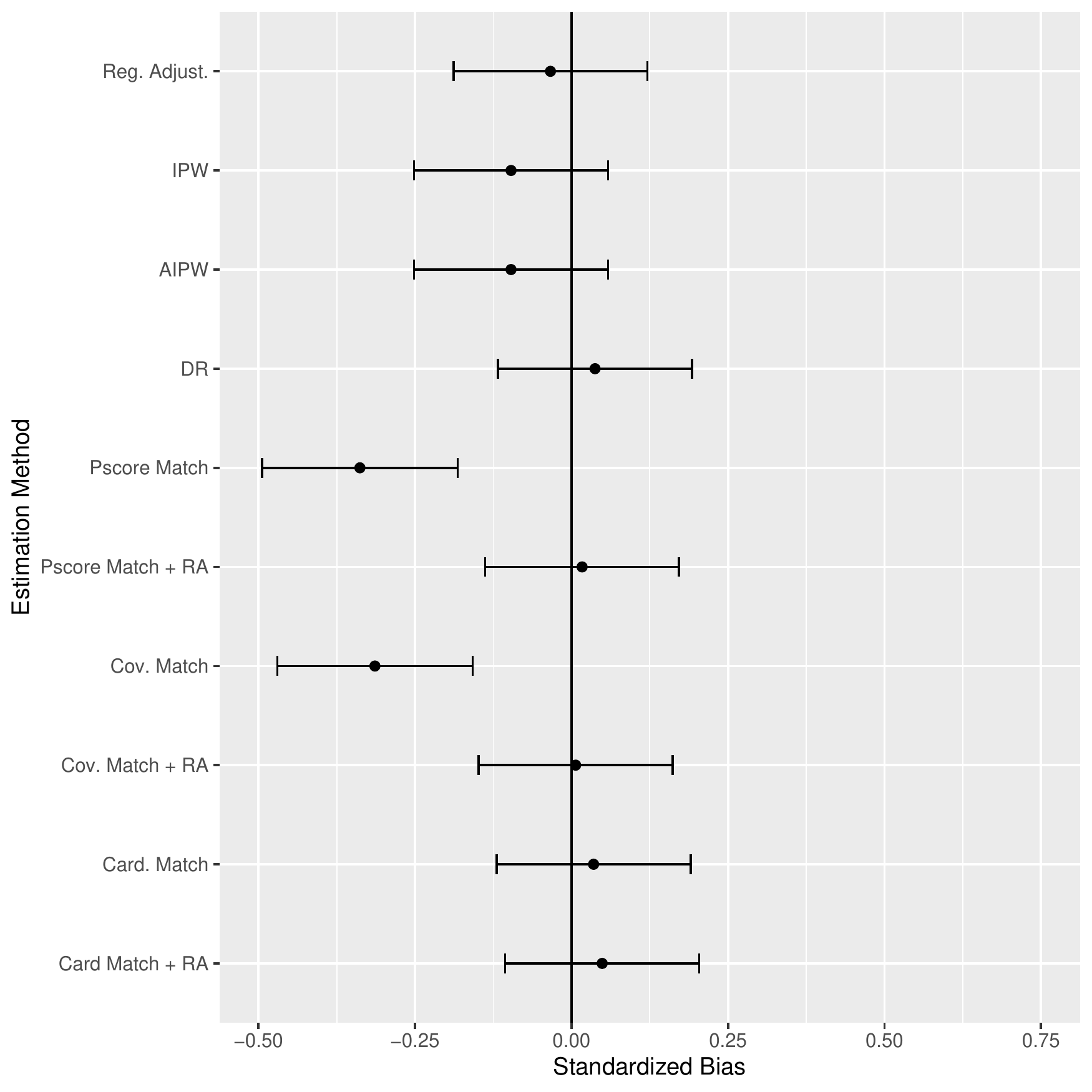}
  \caption{Quality of Life}
  \label{fig:r2}
\end{subfigure}
\caption{Bias in statistical estimates based on methods of statistical adjustment available in R. Point estimates are standardized difference between NRS and RCT estimates.}
\label{fig:r_plots}
\end{figure}

Next, we provide two broad summaries of the results. Thus far, we have only compared methods in terms of bias: the difference between the RCT and NRS treatment effect estimates. Next, we report results using the MSE metric outlined above. Under a MSE criterion, we might expect treatment assignment focused methods to perform worse, since they do not utilize outcome information. We calculated MSE as the length of the 95\% confidence interval added to the bias-term squared. 

Table~\ref{tab.1} contains MSE estimates for each method and both outcomes. For the health status outcome, MSE estimates are mostly similar, and use of outcome information confers little benefit. For example, the propensity score match in R has an MSE value of 0.11, which is identical to the estimate based on a Superlearner and TMLE. The one exception is the propensity score match estimates from Stata. For this method, the MSE is double the other methods. For the quality of life outcome, a few patterns are worth noting. Using regression adjustment in conjunction with matching produces very large gains in MSE. Second, the propensity score match in Stata is again the worst performer.  Finally, estimates based on machine learning methods, are not clearly superior to other methods that use outcome information.

\begin{table}[ht]
\centering
\begin{threeparttable}
\caption{Mean-Squared Error Comparison Across All Methods and Outcomes}
\label{tab.1}
\begin{tabular}{lcc}
 \toprule
 & Health Status & Quality of Life \\ 
\midrule
GRF & 0.12 & 11.36 \\ 
  BART & 0.08 & 17.02 \\ 
  SL + TMLE & 0.11 & 37.26 \\ 
  R -- Regression Adjustment & 0.09 & 9.81 \\ 
  R -- IPW  & 0.13 & 14.71 \\ 
  R -- AIPW  & 0.13 & 14.73 \\ 
  R -- Doubly Robust Weight. & 0.15 & 8.78 \\ 
  R -- Pscore Match  & 0.11 & 71.03 \\ 
  R -- Pscore Match + Reg. Adj. & 0.09 & 8.51 \\ 
  R -- Covariate Match  & 0.10 & 62.49 \\ 
  R -- Covariate Match + Reg. Adj. & 0.10 & 8.60 \\ 
  R -- Cardinality Match  & 0.15 & 10.11 \\ 
  R -- Cardinality Match + Reg. Adj. & 0.11 & 9.47 \\ 
  Stata -- Regression Adjustment & 0.13 & 43.20 \\ 
  Stata -- IPW  & 0.15 & 26.73 \\ 
  Stata -- AIPW  & 0.10 & 9.15 \\ 
  Stata -- IPW + Reg. Adj. & 0.15 & 26.73 \\ 
  Stata -- PS Match  & 0.28 & 174.07 \\ 
  Stata -- NN Match  & 0.10 & 25.63 \\ 
  Stata -- NN Match + Reg. Adj. & 0.09 & 64.81 \\ 
\bottomrule
\end{tabular}
\begin{tablenotes}[para]
Note: Cell entries are estimated MSE for each method of adjustment.
\end{tablenotes}
\end{threeparttable}
\end{table}
 
Table~\ref{tab.2} contains a full summaries of the complete results in terms of bias. In some respects, the lack of pattern is striking.  That is, in each category there are methods that succeeded and failed. However, modeling the outcome and treatment assignment more often resulted in smaller amounts of bias. The other clear pattern is that simple propensity score matching estimators often displayed larger amounts of bias.

\begin{table}[h]
\centering
\begin{threeparttable}
\caption{Complete Summary of Results Across All Methods and Outcomes}
\label{tab.2}
\begin{tabular}{llcc}
 \toprule
Outcome Model & & Health Status & Quality of Life \\
\midrule
 & Stata -- Regression Adjustment & $\bigstar$ & \textbf{X}\\
 & R -- Regression Adjustment & $\bigstar$ & $\bigstar$\\
 & BART & $\bigstar$ & $\bigstar$\\
\midrule
Treatment Model & & & \\ 
\midrule
 & Stata -- IPW  & \textbf{X} & $\bigstar$\\
 & Stata -- AIPW  & $\bigstar$ & $\bigstar$\\
 & Stata -- PS Match  & \textbf{X} & \textbf{X}\\
 & Stata -- NN Match  & \textbf{X} & \textbf{X}\\
 & R -- IPW  & $\bigstar$ & $\bigstar$ \\
 & R -- AIPW  & $\bigstar$ & $\bigstar$ \\
 & R -- Pscore Match  & \textbf{X} & \textbf{X}\\
 & R -- Covariate Match  & $\bigstar$ & \textbf{X} \\
 & R -- Cardinality Match  & $\bigstar$ & $\bigstar$ \\
\midrule
Outcome and Treatment & & & \\
\midrule
 & Stata -- IPW + Reg. Adj. & $\bigstar$ & \textbf{X} \\
 & Stata -- NN Match + Reg. Adj. & $\bigstar$ & \textbf{X} \\
 & R -- Doubly Robust Weight. & $\bigstar$ & $\bigstar$ \\
 & R -- Pscore Match + Reg. Adj. & $\bigstar$ & $\bigstar$ \\
 & R -- Covariate Match + Reg. Adj. & $\bigstar$ & $\bigstar$ \\
 & R -- Cardinality Match + Reg. Adj. & $\bigstar$ & $\bigstar$ \\
 & R -- GRF & $\bigstar$ & $\bigstar$ \\
 & R -- SL + TMLE & $\bigstar$ & \textbf{X} \\
\bottomrule
\end{tabular}
\begin{tablenotes}[para]
Note: A \textbf{X} indicates NRS estimate was significantly biased. A $\bigstar$ indicates NRS estimate did not significantly differ from RCT estimate.
\end{tablenotes}
\end{threeparttable}
\end{table}

\section{Conclusions}

REFLUX used an innovative design to enroll patients in either a randomized or patient preference arm. This study design harmonized key aspects of the study to ensure high levels of comparability between the two arms of the study. In this manuscript, we exploited this design to study methods of statistical adjustment for treatment effect estimation. By harmonizing the design and and analysis, we can isolate whether any bias in the effect estimates is due to the method of statistical adjustment.
 
Next, we review the broad lessons that can be drawn from the statistical analysis. First, ML based methods of statistical adjustment are not fool proof. For the second outcome, two of the ML methods displayed significant or nearly significant levels of bias. Second, we found that simpler matching estimators tended to display significant levels of bias. In both Stata and R, propensity score matching performed poorly in almost every scenario. As we noted above, propensity score based matching methods generally failed to produce acceptable levels of balance. This points to a key weakness in propensity score matching. If the initial match fails to produce acceptable levels of balance, the analyst has few options for improvement. The general advice in the literature is to alter the specification of the propensity score model \cite{lee2017practical,harder2010propensity,leite2016practical,guo2015propensity}. In our experience, this is rarely successful. More advanced matching methods allow the investigator to improve balance directly through propensity score calipers or through balance constraints \cite{Rosenbaum:2010,zubizarreta2012using,Pimentel:2015a,Sekhon:2013}. As we observed, additional regression adjustment reduces this bias, since regression adjusts the treatment effect estimate by the amount of residual imbalance.  Next, propensity score weighting methods in Stata underperformed those in R. For the second, outcome the results from Stata displayed significant levels of bias, while those in R did not. This is due to the fact that the match in Stata is a greedy method, while the match in R was based on an optimal matching method.  Our results agree with recent arguments in the literature against propensity score matching \cite{king2019propensity}. Finally, the more advanced match based on cardinality matching performed well for both outcomes even without additional bias reduction using regression models. This result complements the simulation based results where cardinality matching also tended to outperform other alternatives \cite{resa2016evaluation}.

The variation in results does, however, suggest one important conclusion: analysts should typically estimate effects using more than one method of statistical adjustment. The variation we found was not related to approach or estimation method. As such, it behooves researchers to check that the results are not dependent on the method of statistical adjustment. It is our sense that most studies only use a single method of statistical adjustment, and this leaves study results vulnerable to the possibility that results are statistical method dependent. One option is an approach suggested in \cite{keeleblackbox2019}.  They recommend an analysis plan where the study first employs a method of matching or weighting to prioritize balance according to subject matter expertise.  These results are then checked against estimates from ML methods that do not use prioritization and utilize outcome information. This form of analysis plan builds in variation in the methods of statistical adjustment.

\clearpage

\singlespacing

\bibliographystyle{/Users/ljk20/Dropbox/texmf/bibtex/bst/ieeetr}
\bibliography{\string~/Dropbox/texmf/bibtex/bib/keele_revised2}

\end{document}